\documentclass[aps,pra,twocolumn,amsmath,amssymb]{revtex4}
\usepackage{graphicx,epsfig}
\usepackage{bm}
\usepackage{dcolumn}
\usepackage{xcolor}
\usepackage[utf8]{inputenc}
\inputencoding{latin1}
\inputencoding{utf8}
\usepackage[normalem]{ulem}
\usepackage[breaklinks=true,colorlinks,citecolor=blue,linkcolor=blue,urlcolor=blue]{hyperref}

\def\noi{\noindent}
\def\bc{\begin{center}}
\def\ec{\end{center}}              

\usepackage{filemod}
\usepackage{braket}

\newcommand{\bea}{\begin{equation}}
\newcommand{\eea}{\end{equation}\noi}
\newcommand{\ber}{\begin{eqnarray}}
\newcommand{\eer}{\end{eqnarray}\noi}
\begin{document}
\title{Enhancement in performance of quantum battery by ordered and
disordered interactions }

\author{Srijon Ghosh$^1$, Titas Chanda$^{1,2}$, Aditi Sen(De)$^1$}

\affiliation{\(^1\)Harish-Chandra Research Institute and HBNI, Chhatnag Road, Jhunsi, Allahabad - 211019, India\\
\(^2\)Instytut Fizyki im. Mariana Smoluchowskiego, Uniwersytet Jagiello\'nski,  \L{}ojasiewicza 11, 30-348 Krak\'ow, Poland}

\begin{abstract}

Considering ground state of a quantum spin model as the initial state of the quantum battery, we show that both ordered and disordered interaction strengths play a crucial role to increase the  extraction of power from it. In particular, we demonstrate that exchange interactions in the $xy$-plane and in the $z$-direction, leading to the $XYZ$ spin chain, along with  local charging field in the $x$-direction substantially enhance  the efficiency of the battery  compared to the model without interactions. Moreover, such an advantage in power obtained due to interactions is almost independent of the system size.   We find that the behavior of the power, although measured during dynamics, can faithfully mimic the equilibrium quantum phase transitions present in the model. We observe that with the proper tuning of system parameters,  initial state prepared at finite temperature can generate higher  power in the battery than that obtained with  zero-temperature. Finally, we report that defects or impurities, instead of  reducing the  performance, can create  larger amount of quenched averaged power  in the battery  in comparison with the situation when the  initial state is produced from the spin chain without disorder, thereby showing the disorder-induced order in dynamics. 

\end{abstract}

\maketitle

\section{Introduction}

In modern era, devices which store energy for  later purposes are extremely useful to fulfil our daily needs ranging from communication appliances to  medical accessories like  artificial cardiac pacemakers, hearing aids. Prominent examples of such energy storage  include batteries consisting of one or more chemical or electrochemical cells, converting chemical energy to electrical one. They can  either be disposable or rechargeable -- the later ones can be charged externally by using electricity and are very convenient due to their multiple usage facilities. On the other hand, it has been realized over a last few decades that technologies like computers, communication gadgets based on quantum mechanical principles can perform more efficiently than their classical analogs \cite{NC}. Importantly, such devices have already been built  in laboratories by using physical systems like photons, ion-traps, superconducting qubits \cite{photon, new1, iontraps, new2, new3, new4, coldatoms, new5, new6, superconducting, new7}. 

It is therefore natural to ask whether quantum mechanical properties like  coherence \cite{cohrev}, entanglement \cite{entrev} can also play a role to efficiently store or generate  energy. In this respect,  two distinctly different versions of  quantum  batteries are proposed  -- (1)  arbitrary number of independent quantum systems acts as  cells of a battery and entangling unitary or nonunitary operations  are applied  for a suitable time period to drive the system leading to  the extraction of energy from it \cite{Alicki_Fannes, qB-Alickitype, new8, new9, new10, new11, new12, new13, new15, new16, batteryreview}; (2) secondly,  the ground state of an interacting spin model can be considered as the initial state of the battery which can then be used as a storage media where  charging is performed via quantum mechanically allowed operations \cite{qb_spinchain, new17}. Although the former proposal have extensively been studied in recent years, the later one have recently been explored and was shown that nature of coupling of the initial ordered Hamiltonian is crucial for obtaining the improvement in the power \cite{batteryreview}. In this paper, we concentrate on the second kind where the initial state of the battery is prepared in the ground or thermal state of the quantum spin chain and \emph{local} charging field is used to to drive the system required to extract power from the battery. 
With the development of  ultracold atoms trapped in optical lattices or in trapped ions or in polar molecules, the basic ingredient for quantum battery, quantum many-body Hamiltonians, can currently be implemented  and engineered in laboratories, thereby creating possibilities of manufacturing quantum technologies using these systems \cite{coldatoms, iontraps, amaderadv, polarmoleculesexp, new18, new19}.

 On the other hand, the  systems without any impurities or defects  are in general difficult to build and at the same time, keeping them at absolute zero temperature is also hard. 
Therefore, disordered systems \cite{disorder-original, new20, new21, disorderreview, new22, new23, new24, new25, new26} and effects of temperature on physical properties of many-body systems  have attracted lots of attentions in recent times \cite{Sachdevbook, tempinduced, new27, new28, new29, new30, new31, new32}. Moreover, it was discovered that the  disordered models posses exotic phases like Bose  glass  \cite{Boseglass, 33, 34, 35} (cf. \cite{Fermiglass, disphases, 36, 37, 38})  which are not present in the homogeneous systems as well as can show counter-intuitive phenomena like Anderson localization \cite{disorder-original}, many-body localization \cite{many-bodyL, 39, 40, 41, 42}, high-\(T_c\) superconductivity \cite{high_T_c, 43}.  These disordered systems can  also  be created in a controlled manner in ultracold gases, and hence one can  observe these phenomena and quantum phases   in experiments, making this field more appealing \cite{disorderexp, 44, 45, 46, 47, 48}. 

In this paper, we first investigate the role of many-body interactions, ordered as well as disordered, of the parent Hamiltonian and the temperature of the initial state on the efficiency of the battery.  
Specifically, we show that  in case of  the transverse $XY$ and the $XYZ$ model without disorder, power of the battery critically depends  on the interactions and  its  characteristics   like the  ferromagnetic or  the antiferromagnetic ones. We also find that  the  advantages in power generation due to the interactions remain almost same for different system sizes. 
Moreover, signatures of  quantum critical points, present in these models, are clearly visible in  the trends of the power.   Note that although the output power of the battery is measured in the evolution of the system, it can still indicate the equilibrium property of the parent Hamiltonian (cf. \cite{DQPT, 49}).   We also show that suitable tuning of interactions and temperature lead to a situation where power of the quantum battery increases with the increase of temperature, although  one intuitively  expects that the  initial state prepared at high temperature  can destroy the effectiveness of the quantum battery. Moreover, we observe that the  Gaussian-distributed random interaction  strengths, both in the $xy$-plane and  in the $z$-direction of the $XYZ$ model, enhance the quenched-averaged power compared to that of the ordered case. Such counter-intuitive phenomena were already demonstrated in physical quantities like magnetization,  correlation length, entanglement computed in the static scenario i.e., in   the ground or  in the thermal states of the disordered models \cite{disorderinducedorder, 50, 51, 52, 53, 54, 55, 56, 57, 58, 59, 60, 61, 62, 63, 64, disinducorderent, 65, 66, 67}.  Our results indicate that such  advantages can also be found in closed dynamics of the systems with defects.

The paper is organized as follows: In Sec. \ref{sec-QBset}, we introduce the concept of quantum battery and the respective measure to quantify its efficiency. We then discuss the quantum spin models, both ordered and disordered ones,  that we use for modelling quantum battery (Sec. \ref{sec-qspinmodel}). We then present the results in Sec. \ref{sec-results} for ordered spin models with the initial states of the battery being either the ground state or the thermal state with finite temperature. Finally, we show that models with random exchange interactions can increase the quenched averaged  power of the battery in Sec. \ref{sec-disorder}. The conclusion is in Sec. \ref{sec-conclu}.

%
%
%
%

\section{Quantum Battery built from quantum spin chain: Set the stage}
\label{sec-QBset}

\begin{figure}
\includegraphics[width=\linewidth]{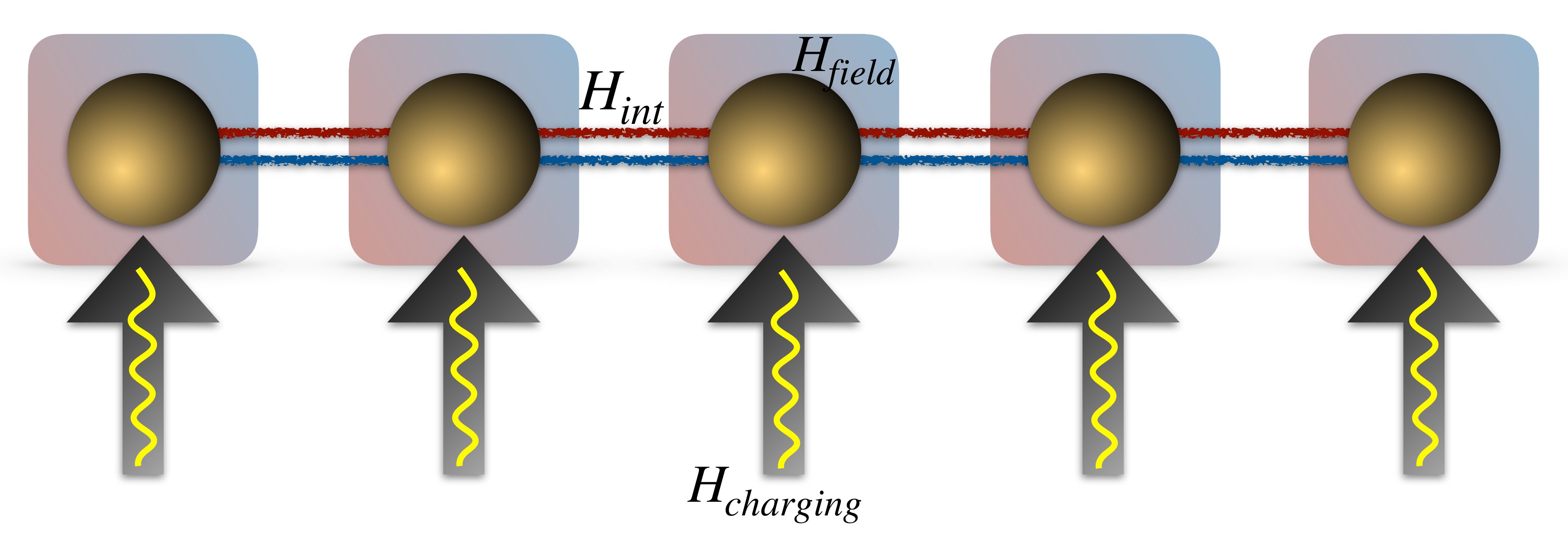} 
\caption{Schematic diagram of a quantum battery. Initially, thermal or the ground state  of a spin chain,  having interaction part,  $H_{int}$ and  the  local magnetic field part, $H_{field}$, acts as a quantum battery. It is then driven by  the local magnetic field, $H_{charging}$ to extract maximal power from the battery. }
\label{fig:schematic}
\end{figure}

A quantum battery is usually considered as  $N$ identical and  independent quantum mechanical systems, in arbitrary dimension, expressed by a Hamiltonian, $H_0$, having non-degenarate eigenvalues. To extract work, the system is driven by an interacting Hamiltonian,  acting on the total $N$-party system, $H_{charging}^g$,  which can, in general,  be time-dependent \cite {Alicki_Fannes, qB-Alickitype,  new8, new9, new10, new11, new12, new13,  new15, new16, batteryreview}. Such Hamiltonian can, in principle, create entanglement in the dynamical state.

 In contrast to this, we choose a quantum battery, made up  of  $N$ interacting spin-$\frac{1}{2}$ particles governed by a  Hamiltonian, \(H_0\).  In this work, one of our primary goal is to study the effect of  interactions and its nature  on the efficiency of the battery. Hence  the Hamiltonian considered here constitutes of two parts,  given by
\begin{equation}
H_0 = H_{field} + H_{int},
\end{equation}
where $H_{field}$ represents the external local magnetic field, while $H_{int}$ is two or more-body  interactions between the spins of the spin-chain. 
To drive the system (or more precisely, the battery), a local charging field $H_{charging}$, is applied on each individual spin. See Fig. \ref{fig:schematic} for the schematic representation of the battery. With $H_{int}=0$, the battery and its charging process only consist  of local terms, so that it becomes exactly analogous to a  situation which cannot posess any quantum features like entanglement or quantum discord. Note that the similar scenario is considered in Ref. \cite{qb_spinchain, new17} although unlike the local field,   the interacting part of the Hamiltonian along with the charging field is employed to extract the work from the battery.

Let us first notice that one can trivially increase the efficiency of the battery by multiplying some constant (greater than one) to $H_0$, or by
increasing the magnitude of the local part, $H_{field}$, of the Hamiltonian. To make the analysis non-trivial,  we normalize $H_0$ as 
\begin{eqnarray}
 \frac{1}{E_{max}-E_{min}}[2H_0-(E_{max}+E_{min}) \mathbb{I}]  \rightarrow H_0
\end{eqnarray}
where $E_{min}$ and $E_{max}$ are minimum and maximum energy eigenvalues of  \(H_0\) respectively. Due to this normalization, the spectrum of $H_0$ is now bounded in $[-1,1]$ irrespective of the parameter values. This normalization enables us to exactly find out the consequence of $H_{int}$ in power compared to the case with vanishing  interaction part, i.e., \(H_{int}=0\) which may not have any quantum characteristics.    

The charging of the battery in  a closed system takes place  according to the unitary operator, given by
\begin{equation}
U(t)= \exp(-iH_{charging}t),
\end{equation}
which is responsible for the time-evolution of the initial state, $\rho(t=0)$, of the battery.  Initially, the battery is prepared either in (i) the ground state of the normalized Hamiltonian, which corresponds to the situation of absolute zero temperature, or in (ii) the canonical equilibrium state, $ \rho_{th}= \exp(-\beta H_0)/{Z}$, for a given inverse temperature, $\beta = 1/k_B T$, with $Z=\mbox{Tr} (\exp (- \beta H_0))$ and $k_B$ being the corresponding  partition function and the Boltzmann constant respectively. It is important to note here that since the absolute zero temperature is hard to achieve in experiment, a state with finite temperature is a natural choice for the initial state of the  battery.
%
%
%
%
At a particular time instant $t$, the total work-output by the battery can be  defined as
\begin{equation}
W(t)= \mbox{Tr}(H_{0}\rho(t))-\mbox{Tr}(H_{0}\rho(t=0)),
\end{equation}
where \(\rho(t)= U(t) \rho(t=0) U(t)^{\dagger}\) is the evolved state of the system.  The corresponding average power for a given time $t$  can  be written as
 $P(t)=\frac{W(t)}{t}$. 
 The aim in preparing the battery is to maximize the extractable power, and hence it is important to choose a proper time when the evolution should be stopped.   
Towards this objective,   the maximum average power obtained from a given battery can be quantified as
\begin{equation}
P_{max}= \max_{t}\frac{W(t)}{t},
\end{equation}
where the maximization is performed over time, \(t\). In the rest of the paper, we call \(P_{max}\) as the power of the battery which is the maximum power, obtained in optimized time. 

We use exact diagonalization techniques to obtain the ground (or thermal) states as well as the evolved states. For optimization over time, $t$, we first use global optimization algorithms (simulated annealing and straightforward grid method), and then employ the widely used COBYLA local optimization algorithm \cite{book2}.

\section{Quantum Spin  model as Battery}
\label{sec-qspinmodel}

Let us  describe the properties of quantum $XYZ$ Heisenberg spin chain with magnetic field which we consider as  $H_0$. Its ground or canonical equilibrium state serves as the  possible initial state of the battery. 
The Hamiltonian consisting of  $N$ spin-$1/2$ particles with open boundary condition reads as
\begin{widetext}
\begin{eqnarray}
H_0= \underbrace{\frac{1}{2} h \sum_{j=1}^N  \sigma_{j}^z}_{H_{field}} + \underbrace{\frac{1}{4}\sum_{j=1}^{N-1} J_{j}[(1+\gamma)\sigma_{j}^x\otimes\sigma_{j+1}^x+(1-\gamma)\sigma_{j}^y\otimes\sigma_{j+1}^y]+\frac{1}{4}\sum_{j=1}^{N-1} \Delta_{j}\sigma_{j}^z\otimes\sigma_{j+1}^z}_{H_{int}},
\label{eq_mainHamil}
\end{eqnarray}
\end{widetext}
where  $\sigma^{\alpha}$ $(\alpha = x,y,z)$ represents the usual Pauli spin matrices,  $h$ is the strength of the external magnetic field at each site, $0 \leq \gamma \leq 1$  is the anisotropy constant, and  $\{J_{j}\}$,  $\{\Delta_{j}\}$ are the nearest neighbor coupling constants in the $xy$-plane and in the $z$-direction respectively.  They may or may not depend on site $j$. 
In a closed system, the quantum battery can  be charged by applying local  external magnetic field in the $x$-direction with strength $\omega$, as
\begin{equation}
H_{charging}=\frac{\omega}{2}\sum_{j=1}^N \sigma_{j}^x.
\label{eq_chargingHamil}
\end{equation}
To obtain the work and then power of the battery, the time-dynamics is computed by constructing the unitary operator via Eq. (\ref{eq_chargingHamil}) where the ground or the thermal state of the spin model in Eq. (\ref{eq_mainHamil}) is used as the initial state.  It is important to stress here that realizability of these models by currently available technologies create possibilities to implement the proposed battery   in laboratories.

\subsection{Quantum XYZ Heisenberg model with homogeneous interaction}

Depending on the scenarios, whether the sets, $\{J_{j}\}$  or   $\{\Delta_{j}\}$ is site-independent or not, the spin-system can be called ordered or disordered. In this paper, we will explore both the cases.
Let us first  consider the system with  $J_{j} = J$ and $\Delta_{j} =\Delta$, i.e. the parameters involved in Eq. (\ref{eq_mainHamil}) are  site independent, leading to the ordered spin chain.  In one dimension, Eq. (\ref{eq_mainHamil}) represents a paradigmatic families of Hamiltonians with nearest neighbor interactions, having a rich phase diagram at zero temperature. 
%
Let us now discuss some important sub-classes of \(H_0\),
and their phase portraits.  
 
\begin{enumerate}

\item  $\Delta=0$,  and $\gamma \geq 0$ \cite{LSM, Barouch-McCoy, 68}:  \(\gamma =0\)  represents the transverse $XX$ spin chain, while  the $XY$ spin model  having transverse magnetic field is with $\gamma \neq 0$. They belong to two different universality  classes -- the former one has a \emph{gapless} spin-liquid (SL) phase for $|J/h| > 1$, and a paramagnetic (PM) phase for $|J/h| < 1$, 
while the later one belongs to the Ising universality class, consisting of a PM  $(|J/h| < 1)$, an  antiferromagnetic (AFM)  $(J/|h| > 1)$, and a ferromagnetic (FM)  $(J/|h| < -1)$ phases.  Both the models can be solved analytically by Jordan-Wigner transformations \cite{LSM, Barouch-McCoy, 68} for arbitrary system size including in the thermodynamic limit.

\item $\gamma=0$,  $\Delta \neq 0$ \cite{Takahashibook, muller85, Dimitriev02}: The model  is known as the $XXZ$ spin chain. For $h=0$, the model is integrable -- with $J=1$, there is an AFM region for $\Delta > 1$, and  $\Delta < -1$ corresponds to the ferromagnetic (FM) one, while $-1 < \Delta < 1$ is the gapless SL phase.  By using different approximate and numerical techniques,  quantum critical lines and their corresponding phases of  the system with $h \neq 0 $ has also been explored \cite{Dimitriev02}. For example, with small values of magnetic field and \(\Delta\), a new phase, N\'eel order in the $y$-direction, develops, which is known  as spin-flop phase (SF). 

\item   $\Delta  \neq 0$, $\gamma \geq 0$ ($XYZ$ model)\cite{BakROPP, Sela11, realizationXYZproposal}:   The model is not exactly solvable. Several numerical and approximate studies of the $XYZ$ model with field reveal that it has a very rich phase diagram. In particular, like the $XXZ$ model, it also posses FM, AFM and SF phases although for non-zero values of \(\gamma\),  two new quantum phase transitions \cite{KTtransition, 69} of different kinds appear -- one from SF to a new phase called gapless floating phase (FP), while another one is from the FP to the AFM phase. 
\end{enumerate} 
\noindent We will show in the next section that tuning parameters leading to different quantum spin models play an essential role to  build and maintain the performance of the battery. 


\subsection{Quantum $XYZ$ model with random interaction strength: Disordered quantum spin model }
\label{subsec_disorderQXYZ}

Let us now consider the system, in which one of the interaction strengths are chosen randomly. It can be found during the preparation process of the materials or due to dislocations of atoms from their regular lattice sites or due to environmental effects \cite{Guo, Rieger, Rieger1, Vojta, book1}. Since the change of disorder in these systems remains almost fixed for certain times, specifically a much longer duration than that of the evolution of the system, this kind of disorder can be called ``quenched" which we will consider in this paper. It can also  be created and controlled  in laboratories with cold atoms in optical lattices, linear chains of ions etc. \cite{disorderexp, 44, 45, 46, 47, 48}.
In this paper, two situations are  considered which are as follows:

\begin{enumerate}

\item  The nearest neighbor exchange interaction in the $xy$-plane, $\{J_{j}/|h|\}$, are randomly chosen  from a Gaussian distribution with mean $\overline{J}/|h|$ and the standard deviation $\sigma_J$ which we refer as the strength of disorder. \(\sigma_J =0\) corresponds to the ordered case.  Here,  $\{\Delta_{j} /|h|\}=\Delta/|h|$ remains independent  of the sites. Quenched averaging is performed by first computing the power of the battery  for each realization with random-distributed   $\{J_{j}/|h|\}$,  and then by taking the average over all realizations.  Mathematically, for a physical quantity, $\mathcal{O}$,  and for a randomly chosen parameter, $\{X_{j}\}$, with mean \(\overline{X}\) and standard deviation \(\sigma_X\) involved in the system, quenched averaged quantity can be represented as 
\begin{equation}
\langle O  ( \overline{X}, \sigma_X) \rangle  = \int\int...\int \mathcal{O} \lbrace X_{j} \rbrace d \lbrace X_{j} \rbrace, 
\end{equation} 
where the integration is carried out with respect to the probability distribution by which the  $\{X_{j}\}$ are chosen. 
In our case, the power of the quantum battery ($P_{max}$) is the physical quantity, which has to be quenched averaged over the parameter-space, $\{J_{j}/|h|\}$, denoted by $\langle P_{max} \rangle$.  


\item Fixing $\{J_{j}/|h|\}= J/|h|, \forall j$,  we also study the effect of disorder on power by choosing  $\{\Delta_{j} / |h|\}$ randomly from  a Gaussian distribution with mean  $\overline{\Delta}/|h|$ and standard deviation $\sigma_{\Delta}$.

\end{enumerate}

\begin{figure}
\includegraphics[width=0.6\linewidth]{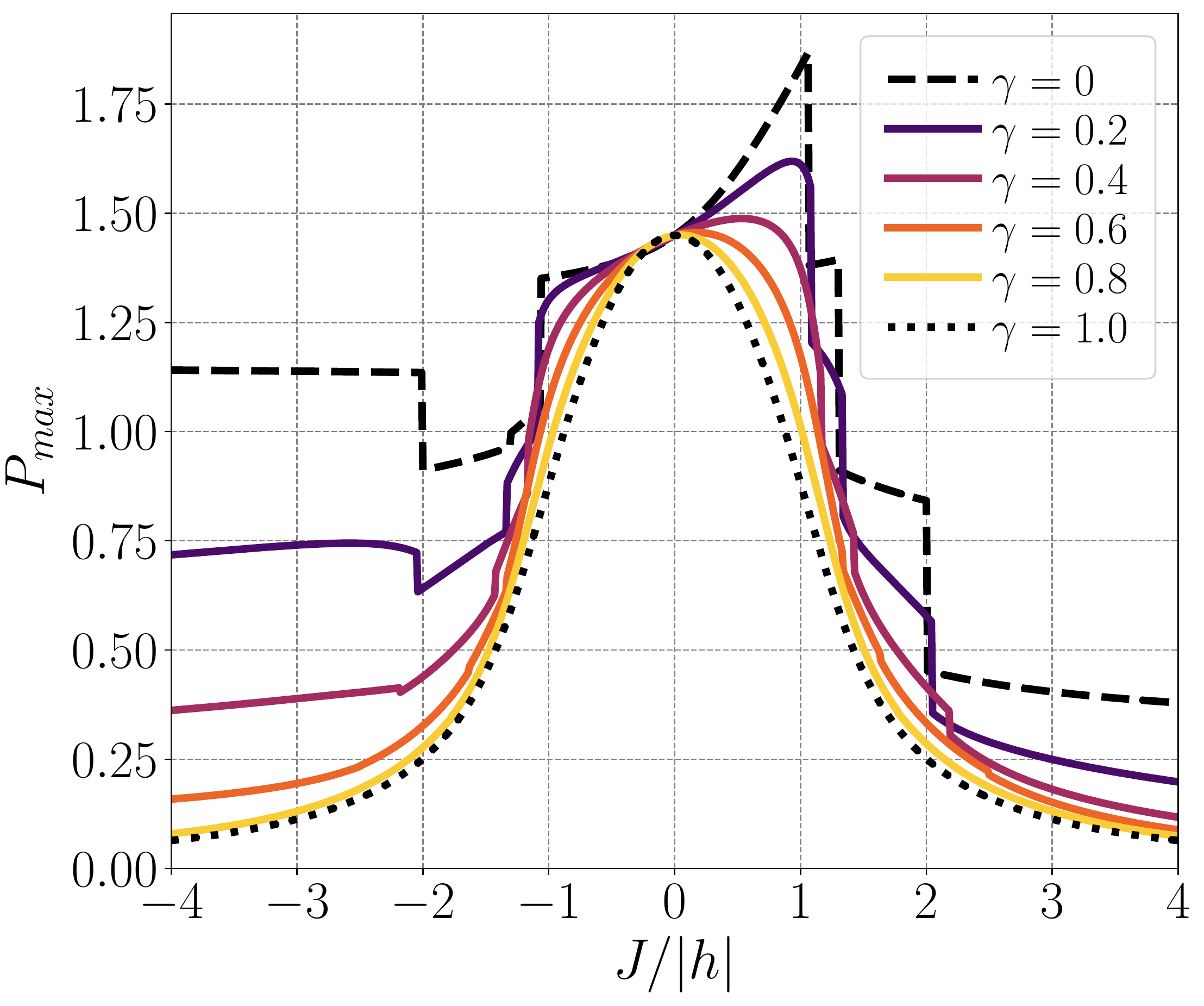}
\caption{(Color online.)  $P_{max}$ (ordinate) vs. $J/|h|$ (abcissa). 
\(P_{max}\)  is computed for  the transverse  $XY$ model (i.e.,  $\Delta/|h|= 0$) with different values of the anisotropy parameter, $\gamma$. Here $N=8$. In the paper, all the plots are for the same system size, unless mentioned otherwise. Both the axes are dimensionless.  
 }
\label{Fig:powervsJfordifferentgamma;N=8} 
\end{figure}

\section{Interaction enhances the Power: Ordered Case}
\label{sec-results}

In this section, we address the question whether nearest neighbor interactions can be beneficial for increasing the extraction of power from the battery. To demonstrate this, we first consider the ground state as the initial state of the quantum ordered $XY$ model with transverse magnetic field as the battery, and then move on to the role of interactions in the $z$-direction  by considering the $XYZ$ model. We further study the effects of finite temperature   on  the efficiency.

\subsection{Effects of interaction term in the $XY$ model}

Let us consider the ground state of the transverse  $XY$ model, and compute the power, $P_{max}$, with the variation of $J/|h|$ for fixed values of system size, $N$.  
The behavior of power, depicted in Fig. \ref{Fig:powervsJfordifferentgamma;N=8}(a),  shows that  the battery prepared by using interacting Hamiltonian has  higher power as output for certain system parameters than that of the system without interactions. For demonstration, we fix some values of $\gamma$, and  the strength of the charging field as \(\omega =2 |h| \).  The interesting observations in the pattern of \(P_{max}\)  are listed below:


%
%


\begin{enumerate}

\item \emph{Positive vs. negative interaction strength. } 
 Positive and negative coupling constants, i.e., $J/|h|>0$ and $J/|h|<0$ indicate the nature of interaction to be antiferomagnetic (AFM) and ferromagnetic (FM) ones.  
As depicted in Fig. \ref{Fig:powervsJfordifferentgamma;N=8}(a), we observe that $P_{max}$ increases when $0 < J/|h| \lesssim 1$,  and reaches its maximum value close to \(J/|h| \approx 1\), while it decreases for  $J/|h| <0$. 
 Typically, static physical quantities,  like magnetization, classical correlators, entanglement \cite{entrev}, in the ground state are symmetric across \(J/|h|=0\)-line \cite{amaderadv, new19, entpapers}. The asymmetry observed here arises due to the choice of uniform charging field in the $x$-direction, given in Eq. (\ref{eq_chargingHamil}) and also the battery Hamiltonian, \(H_0\). Specifically, when interaction strength is large i.e., $|J/h| > 1$, the initial state is either in the AFM phase or in the FM phase where spins are oriented in the $x$-direction for higher values of $\gamma$. Now, since the charging field is in the $x$ direction, it can easily drive the system without demanding more energy, leading to low amount of power generation. On the other hand, when $|J/h| < 1$ i.e. in the PM phase, spins have affinity towards the $z$-direction due to the external field. Therefore, the charging Hamiltonian needs more energy to drive the system out-of-equilibrium, thereby leading to high amount of  power in this phase.
 However, the pattern of $P_{max}\) clearly establishes that the interaction of \(H_0\) helps to improve the performance of the battery in the paramagnetic phase of the $XY$ model. It is clear from the Fig. \ref{Fig:powermaxvsgamma;N=8}(b) that for any values of anisotropy parameter ( \(0 \leq \gamma <1\)), power gets increased in presence of interaction in the PM phase. So, in terms of the enhancement of power we find that the observation is independent of  the anisotropy parameter ($\gamma$).

\begin{figure}
\includegraphics[width=\linewidth]{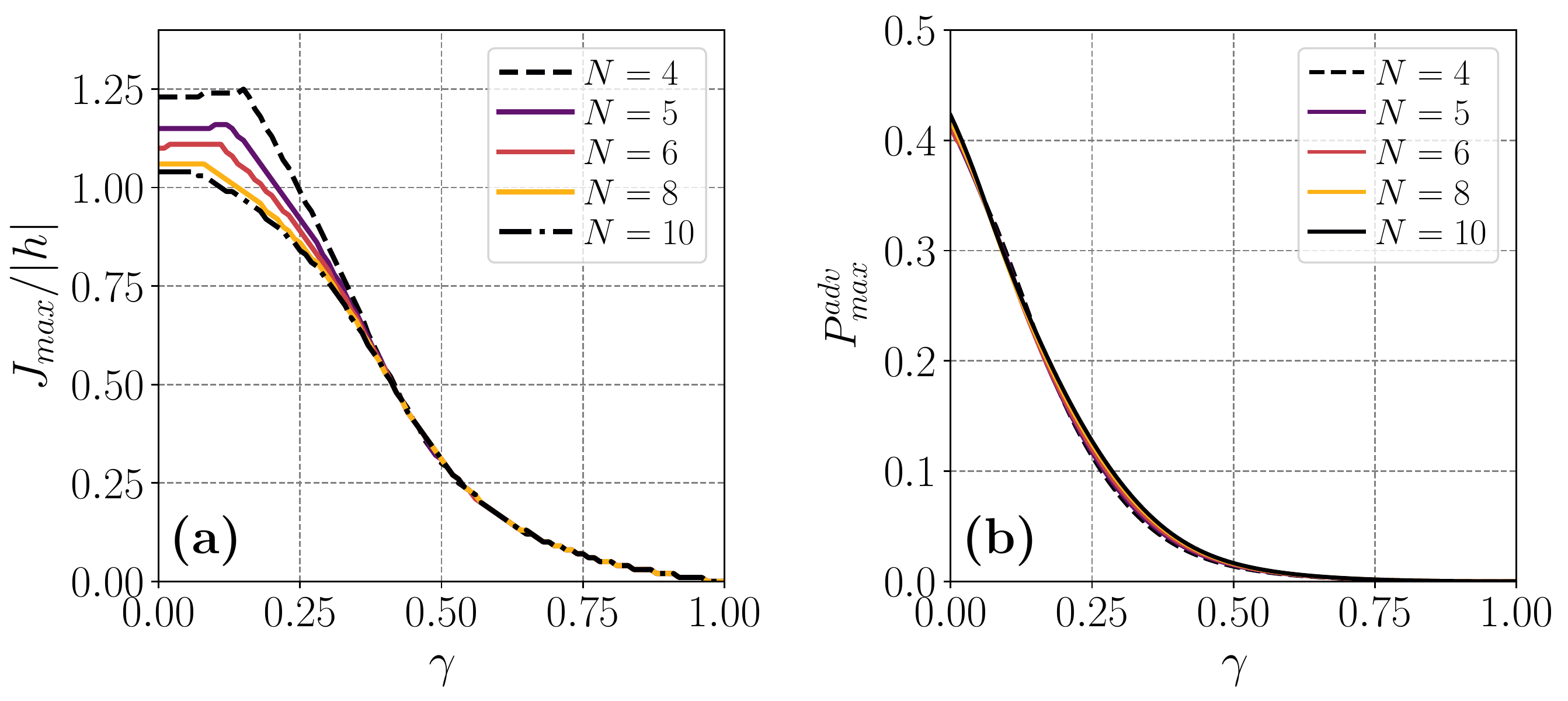}
\caption{(Color online.) Dependence of $P_{max}$ on the interaction strength and the anisotropy parameter, $\gamma$.   Plots are for different system sizes and $\Delta=0$. (a) \(J_{max}/|h|\) vs. \(\gamma\). \( J_{max}/|h| \)  represents the interaction strength for which $P_{max}$ reaches its maximum value  for a given value of  $\gamma$ and  $N$. Note that for higher values of \(\gamma\), \(J_{max}/|h|\) does not depend on \(N\).  (b) \(P_{\max}^{adv}\) against \(\gamma\). The advantages in power due the introduction of $XY$-exchange couplings are measured by the quantity $P_{max}^{adv} = P_{\max}(J_{max}/|h|) - P_{max}(J/|h|=0)$.   Interestingly, \(P_{max}^{adv}\)  becomes scale invariant for the entire range of  $\gamma$.  Both the axes are dimensionless. }
\label{Fig:powermaxvsgamma;N=8} 
\end{figure}

\item \emph{Dependence on $\gamma$.}  Although, the increment of power of the battery is independent of the anisotropy parameter, the magnitude of enhancement, however depends on $\gamma$.  Precisely, maximal power of the battery greatly depends on the anisotropy parameter, as it is evident from  Fig. \ref{Fig:powervsJfordifferentgamma;N=8}(a). Among all the $\gamma$ values,  if the battery is initially in the ground state of the  $XX$ model having  $\gamma =0\), 
the power output is maximum, as compared to  the other values of $\gamma$. 
Also, from Fig. \ref{Fig:powervsJfordifferentgamma;N=8}(a), we find that the range of $J/|h|$, where the advantage in power can be obtained, shrinks with increasing $\gamma$. The reason behind this feature is the same as stated in the previous point, that, with increasing $\gamma$, the strength of exchange interaction in the $y$-direction decreases, and as a result, the tendency to align (or anti-align) in the $y$-direction also decreases. Therefore, it continuously becomes easier for the charging Hamiltonian to drive the system.
To visualize the \(\gamma\) - dependence, we identify the interaction strength,  $J/|h|$, for which \(P_{\max}\) reaches its maximum value, which we refer  as \(J_{\max}/|h|\). We then investigate the behavior of   \(J_{max}/|h|\) with \(\gamma\) for different system sizes, as shown in Fig. \ref{Fig:powermaxvsgamma;N=8}(a).

\begin{figure}
\includegraphics[width=\linewidth]{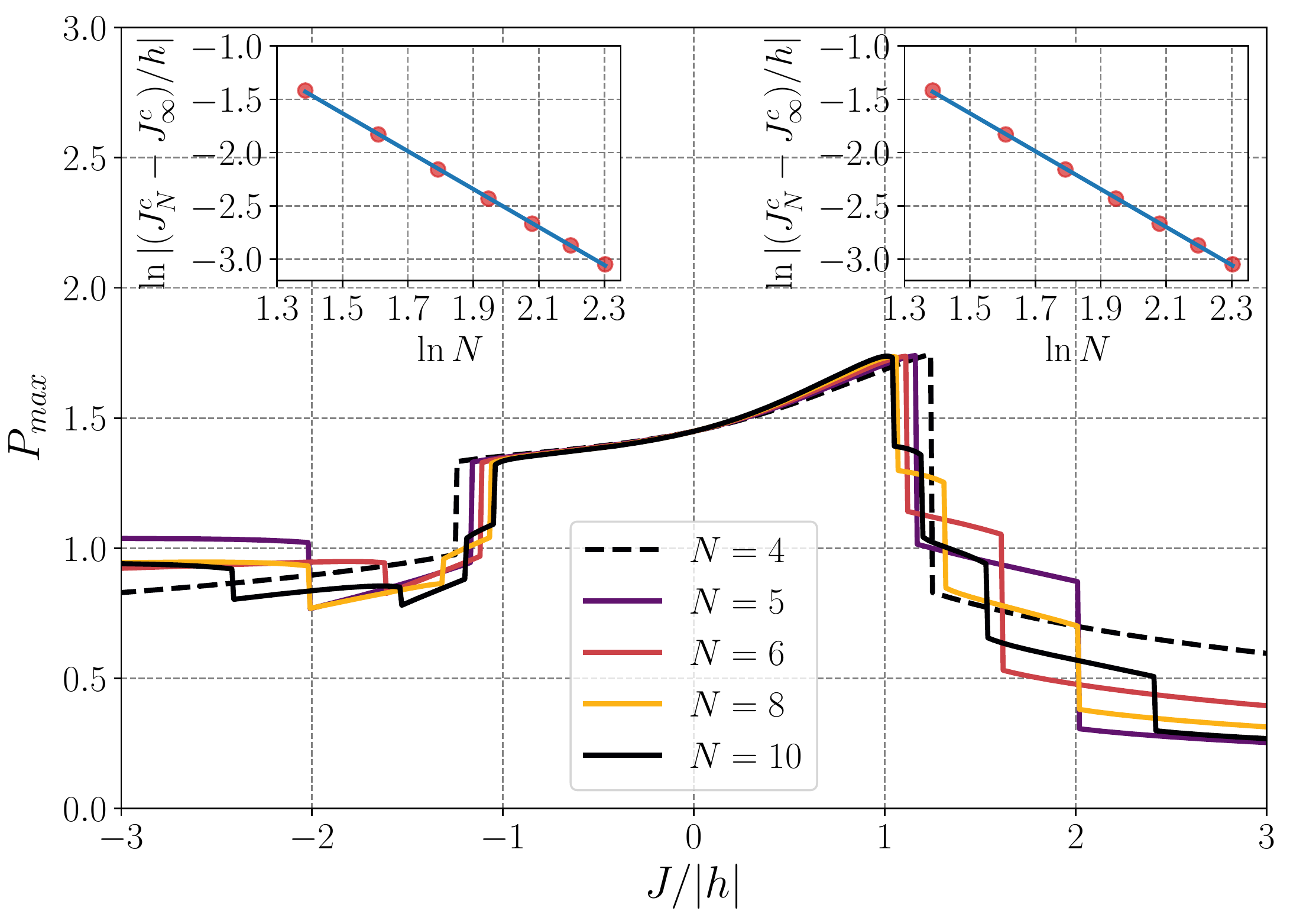} 
\caption{(Color online.) Dependence of power on system size, $N$, for $\gamma=0.1$ and $\Delta=0$. We plot the variation of $P_{max}$ with $J/|h|$ for different system size, $N$. (Insets) Finite-size scaling of the critical points, $J^c_N/|h|$, as indicated by the power. We plot $\ln |(J^c_N-J^c_{\infty})/h|$ (both numerical data and fitted lines) as functions of $\ln N$ for FM $\leftrightarrow$ PM (left inset) and AFM $\leftrightarrow$ PM (right inset) transitions. Both the axes are dimensionless. 
}
\label{Fig:scaling}
\end{figure} 

\item \emph{Role of exchange interaction: Scale invariance.}   The interaction part, \(H_{int}\), in \(H_0\)  is important in \(P_{max}\) as already discussed. To quantify its influence, we introduce a quantity, 
\begin{equation}
P_{max}^{adv} = P_{max}(J_{max}/|h|) - P_{max}(J/|h|=0),
\end{equation}
where \(P_{max}(J_{max}/|h|)\) and \( P_{max}(J/|h|=0)\) are respectively power measured at \(J_{max}/|h|\) defined above and at \( J/|h| =0\).  
\(P_{max}^{adv}\) reaches its maximum value at \(\gamma=0\), and decreases with the increase of \(\gamma\) as seen 
 in Fig.  \ref{Fig:powermaxvsgamma;N=8}(b). Specifically, we find that when \(\gamma =0\),  nonvanishing interaction, in the  chain of $N=8$ sites, can produce  upto \(28.8\%\) increase in power, thereby showing the relevance of quantum battery. Note, however, that  for $\gamma=1$, we find that \(P_{max}^{adv} =0\), i.e., the local scenario is most efficient, and interaction does not help.
Importantly, we observe that \(P_{max}^{adv}\) does not depend on the number of spins in the chain, showing  \emph{scale invariance} property of the advantage.  
 


\item \emph{Quantum phase transition signaled through power.} The second-order quantum phase transition \cite{Sachdevbook, qpt-book, 70} in the $XY$ model  at zero temperature can be detected by the first derivatives  of several physical quantities, which include correlation length \cite{Sachdevbook}, entanglement \cite{entrev}, quantum discord \cite{discord-review, 71} etc.  Since $P_{max}\)  is measured in the evolution, it is not apriori clear that it can identify quantum phase transitions. We here show that for low values of  \(\gamma\), the dynamical quantity,  $P_{max}\) itself, can signal quantum phase transition by showing a finite jump around \(|J/h| \approx 1\).  For higher values of \(\gamma\), \(P_{max}\) changes its curvature from concave to convex so that its derivative shows the kink. It is interesting to note here that in a different context of dynamical phase transition \cite{DQPT, 49}, quantity like Loschmidt echo defined as the distance between the ground and the evolved states of the quantum spin model can also mimic the equilibrium phase transition. Our results, therefore, suggests that  it will be interesting to find (some) other dynamical quantities, similar to power output,  which can also carry the information about the equilibrium phases of the initial systems.

\item \emph{Dependence of power on N. } With the variation of $N$, we observe that in the range of  $-1 \lesssim J/|h| \lesssim 1$,  the  power does not change  its behavior substantially. However,  \(J_{max}/|h|\) which leads to maximum  $P_{max}\) shifts towards \(J/|h| =1\) with the increase of $N$, although the value of the maximum power, as well as maximum advantage in power remain almost unaltered with $N$ (see Figs. \ref{Fig:powermaxvsgamma;N=8} and \ref{Fig:scaling}). This is possible because the curvature of $P_{max}$  becomes  steeper with $N$. On the other hand,  finite-size effects on $P_{max}\) are visible for  \(J/|h| < -1\) as well as  for \(J/|h| >1\) (Figs. \ref{Fig:powervsJfordifferentgamma;N=8}(a) and \ref{Fig:scaling}). 




\item \emph{Scaling.}  Since power of the battery can detect equilibrium quantum phase transition as discussed above, it is now natural to ask the scaling law followed by it. 
Ambitiously,  we find the finite-size scaling of critical points, as indicated by the behavior of  $P_{max}\)
as 
\begin{eqnarray}
\Big|\frac{(J^c_N -  J^c_{\infty})}{ h}\Big| = 1.039 \times N^{-1.78}, 
\end{eqnarray}
for both FM $\leftrightarrow$ PM and AFM $\leftrightarrow$ PM transitions for $\gamma=0.1$ ( Fig. \ref{Fig:scaling}(insets)). Here \(J_N^c/|h|\) is computed where the power shows a first jump for a fixed  value of $N$, while \(J^c_\infty/|h| =1\) as known for the transverse quantum $XY$ model in the thermodynamic limit.  The reason for choosing the first jump in the evaluation of scaling is discussed  in Appendix A. Note, moreover, that we possibly  should not compare the scaling exponent obtained above with the other indicators of QPT -- (1) other physical quantities detecting QPT are calculated in the ground states while the power output is found in dynamics;  the above study establishes that even the dynamical quantity can also carry information about QPT, (2) secondly, the system sizes simulated are too small to demand any comparison.

\end{enumerate}

\subsection{Role of entanglement}

\begin{figure}
\includegraphics[width = \linewidth]{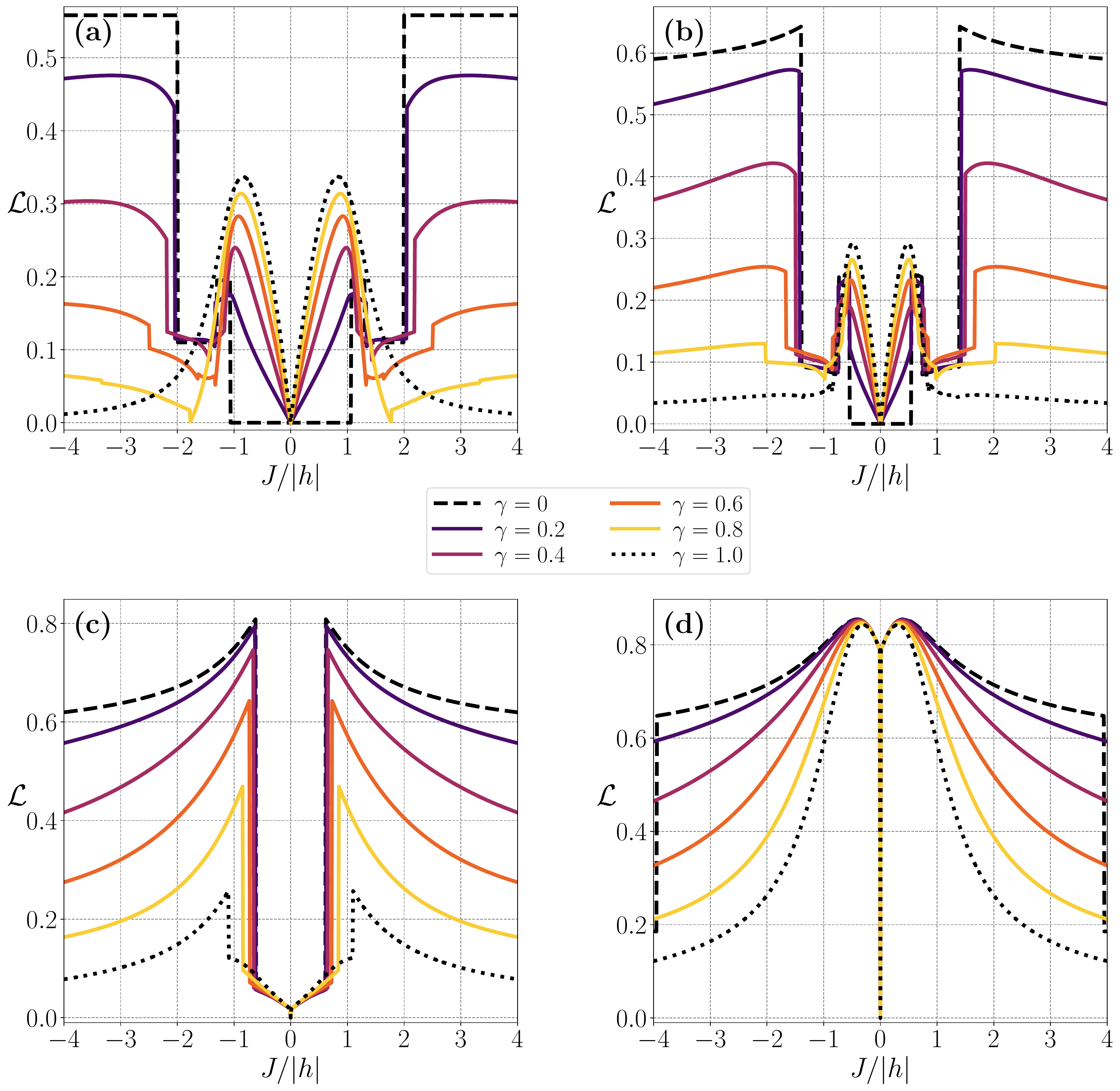}
\caption{(Color online) Plot of nearest neighbor entanglement  of  the ground state in the $XYZ$ model  having  different values of $\gamma$ with the variation of $J/|h|$ ($x$-axis).  Plots are for specific values of  the interaction strengths in the $z$-direction, $\Delta/|h|$, as mentioned in the headings of each plot.  We observe that entanglement is symmetric with \(J/|h| =0\) which is not the case for \(P_{\max}\) (comparing Figs. \ref{Fig:powervsJfordifferentgamma;N=8} and \ref{Fig:powervsJfordifferentgamma_new;N=8}) although the patterns for both of them are qualitatively similar in some regions for \(\gamma >0\). Hence one can argue that entanglement can be a necessary ingredient for good quantum battery, but not sufficient. }
\label{entanglement}  
\end{figure}

We have already shown that many-body interactions can increase the efficiency of a quantum battery. Let us now ask a natural question -- \emph{does inter-spin entanglement play any role in the performance of the battery?} To answer this query, we compute bipartite entanglement \cite{entrev} of the reduced density matrix  obtained by tracing out all the parties except two from the middle of the chain of  both the initial state and the state at the time when $P_{max}$ is optimized. We take the pair of spins from the middle of the chain to minimize any boundary effects due to the open boundary condition. In particular, we calculate logarithmic negativity \cite{72} which is the modulus of the negative eigenvalue of the partial transposed state for two spin-1/2 particles \cite{Peres, Horodecki}.

Let us first consider the XY model. If one compares Figs. \ref{entanglement} (a) and  \ref{Fig:powervsJfordifferentgamma;N=8} with $J/|h| > 0$, we  find that  the nearest neighbor entanglement qualitatively mimics the features of $P_{max}$ -- it increases in  regions, $ -1 \lesssim J/|h|<0$  as well as $ 0 \lesssim J/|h|<1$ and then decreases with \(J/|h|\) for different values of \(\gamma \ne 0\). 
Such characteristics indicates that entanglement can be a necessary ingredient to extract more power, but not sufficient, which is in parity with the earlier results (see \cite{batteryreview} and references therein).

\subsection{Introduction of interaction  in $z$-direction leads to enhancement in Power}

\begin{figure*}[ht]
\includegraphics[width=0.7\linewidth]{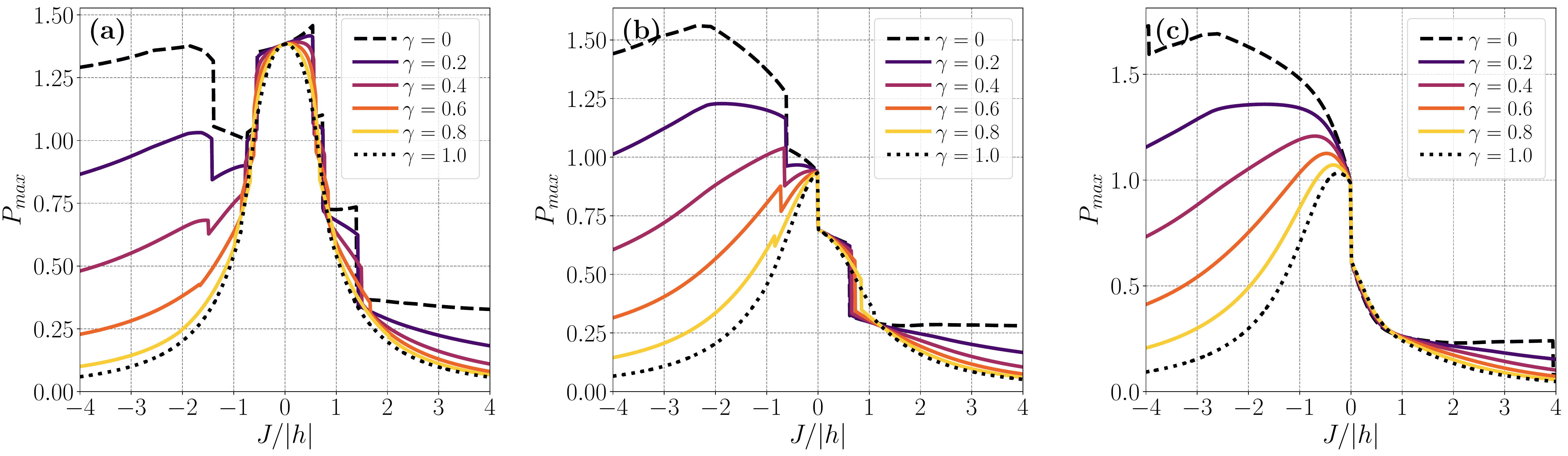}
\caption{(Color online.)  $P_{max}$ (vertical)  agianst $J/|h|$ (horizontal) for the quantum $XYZ$ Heisenberg model with different values of $\gamma$.  
Plots are for specific values of  the interaction strengths in the $z$-direction, $\Delta/|h|$, as mentioned in the headings of each plot. Both the axes are dimensionless.
 }
\label{Fig:powervsJfordifferentgamma_new;N=8} 
\end{figure*}

Let us now move to the $XYZ$ model with magnetic field, given in Eq. (\ref{eq_mainHamil}). We will now address the question whether the additional interactions in the $z$-direction, i.e., the model with \(\Delta/|h| \neq 0\)  is required  to increase  the power of the battery. As before, the battery is initially prepared as  the ground state of this model.

Comparing Figs. \ref{Fig:powervsJfordifferentgamma_new;N=8} with \ref{Fig:powervsJfordifferentgamma;N=8}, we find  that with the increase of \(\Delta/|h|\), the power increases in the region of  \(J/|h| <0\) where the power was decreasing in absence of \(\Delta/|h|\), thereby establishing the usefulness  of  the coupling in the $z$-direction. 
Moreover, we observe that for moderate values of \(\Delta/|h|\),  there is a lower bound on the coupling constant in the $xy$-plane, denoted by \(J_c/|h|  < 0\), where $P_{max}\) increases beyond the value obtained with the initial state of the battery being the ground state of the Hamiltonian without any $XY$ exchange interaction, i.e., with \(J/|h| =0\). Note, however, that the model with \(J/|h|=0\) and \(\Delta/|h|\neq 0\)  corresponds to the  system having nonvanishing interactions, since the field, given to drive the system,  is in the complementary direction of the exchange interaction of the parent Hamiltonian. Again, with the increase of \(\gamma\), \(J_c/|h|\) decreases although  it  is much bigger than that obtained for the $XY$ model. It shows that even if the tuning of the system parameters cannot be performed properly,   the $XYZ$ model  is more appropriate to build the quantum battery than the $XY$ model. 

Although the $XYZ$ model has several competing factors which lead to the generation of high power from the battery, there can be a physical explanation in the same line discussed for the $XY$ model. 
With non-zero $\Delta/|h|$, spins gain another competing tendency which is to anti-align themselves in the $z$-direction. In the PM phase of the XY model with $J/|h| <0\), we do not find any enhancement. However, with the introduction of \(\Delta/|h|\) along with  the field in the $z$-direction,   the charging field in the  $x$-direction possibly requires more energy to drive the system out-of equilibrium, resulting more power. 

\subsection{Effect of temperature on Power of the battery}
\label{sec:thermal}

We have already shown that the zero-temperature state as the  initial state of an interacting Hamiltonian is advantageous for generating high amount of power in the quantum battery. We will now see whether such improvement  persists (or even increases) when the initial state is the thermal state, \(\rho_{th}\), having a finite temperature. This is important because in the laboratory, absolute zero temperature is not easy to obtain. 
To produce power,   local charging Hamiltonian, in Eq. (\ref{eq_chargingHamil}), is again applied to each site. 
 As one  expects, we see that $P_{max}\) vanishes for infinite temperature, i.e.,  for \(\beta =0\),  then starts increasing as \(\beta\) increases, and finally saturates to the power of the zero-temperature. 
However, we notice that the variation of $P_{max}$ with increasing $\beta$ is not always monotonic, and can have one or more \emph{nonmonotonic bumps} depending on the system parameter, which signify that we can have situations, where the battery performs more efficiently at higher temperature than the lower ones. More interestingly, and quite counter-intuitively,  it turns out that battery may output more power at finite temperature than that of the absolute zero temperature. 

\begin{figure}[ht]
\includegraphics[width=\linewidth]{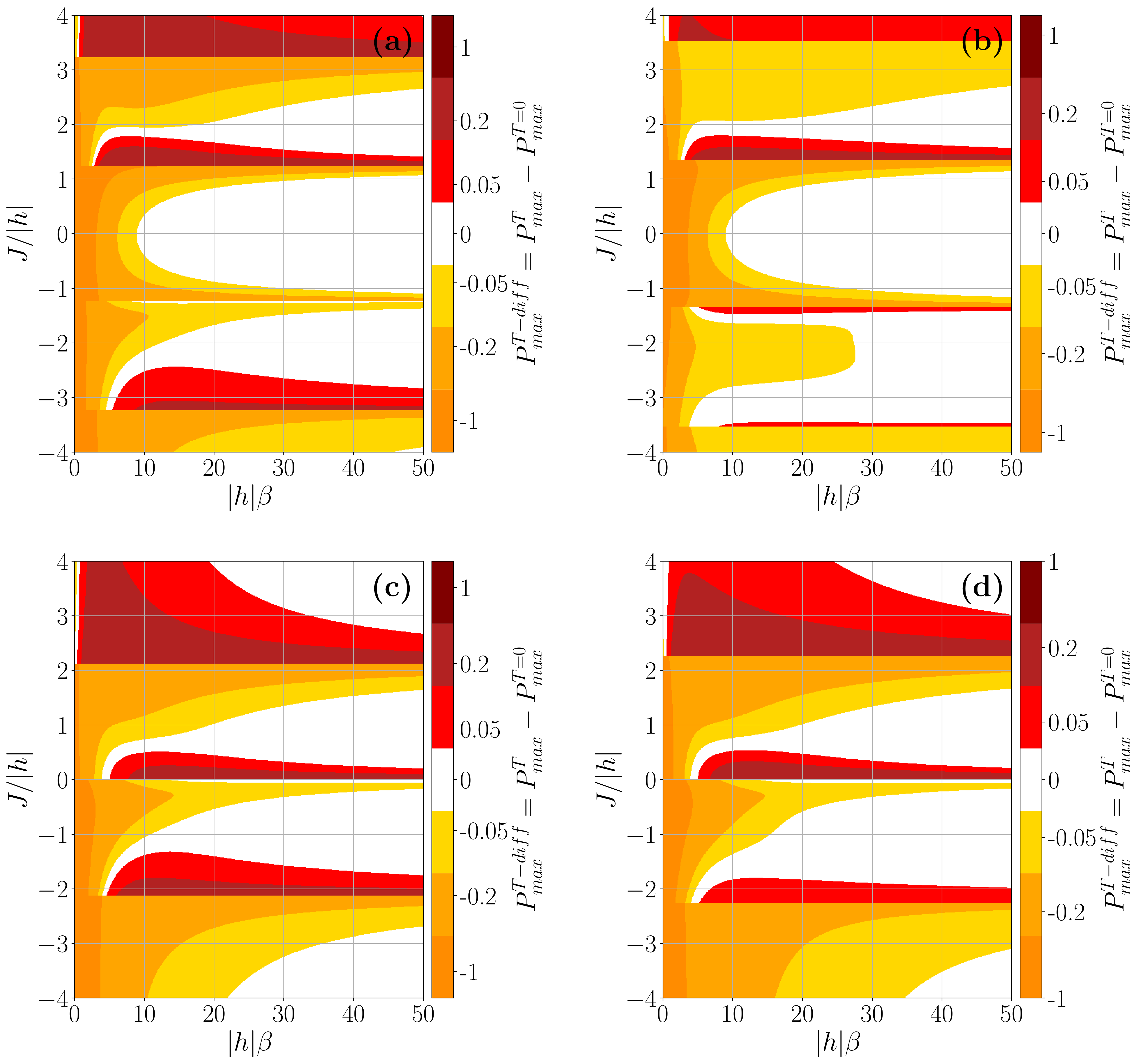} 
\caption{(Color online.)  Map of  $P_{max}^{T-diff}$ (see Sec. \ref{sec:thermal} for definition)
with respect to \(|h|\beta\) (abcissa) and \(J/|h|\) (ordinate). Here $N =4$. (a)-(b): They are for the $XY$ model with two different values of \(\gamma\), \(\gamma =0\) and \(\gamma =0.4\) while (c) and (d) depict the behavior for the $XYZ$ model with \(\Delta/|h| =1\) and the same values of \(\gamma\) as in (a) and (b). 
Positivity of   $P_{max}^{T-diff}$  indicates the advantage of considering initial state at finite temperature while the negative values of  $P_{max}^{T-diff}$ show the benefit for the ground states.
Both the axes are dimensionless. 
 }
\label{fig:therm2}
\end{figure}

Quantitatively, we consider a  quantity  which can capture the advantages gained at finite temperature over the zero-temperature, given by 
\begin{equation}
P_{max}^{T-diff} =P_{max} (T > 0) - P_{max} (T =0),
\end{equation}
where \(P_{max} (T>0)\)  and \(P_{max} (T =0)\) are  the extractable power, obtained with the thermal state and with  the ground state respectively. Indeed, we find that  $P_{max}^{T-diff}$  is positive for  certain choices of  $J/|h|$ and  $\beta/|h|$ (see Fig. \ref{fig:therm2} for four sets of  values of \((\Delta/|h|, \gamma)\)), thereby showing the gain of choosing the thermal state as an initial state.  Numerical simulations also confirm that changing system parameters  does not alter the results qualitatively.

%
%
%

\section{Disorder-enhanced Power from the Battery}
\label{sec-disorder}

In this section, we examine how the presence of impurities  in interactions can induce in  power generation by the battery.  The observations are mainly classified into two situations -- {\bf (i)} random  $XY$ exchange interactions, i.e., randomly chosen Gaussian-distributed $\{J_j/|h|\}$, keeping $\{\Delta_{j}/|h|\} =\Delta/|h|$ fixed for all sites, and {\bf (ii)} disorder in $\{\Delta_{j}/|h|\}$, with  $\{J_j/|h|\}=J/|h|$ being site independent. In general, impurities reduce the physical properties like magnetization, conductivity in systems \cite{disorder-original, new20, new21} and hence the performance of the tasks. However,  
we report that  both the disordered cases considered here can deliver some  advantages  --
(1)  disorder enhances power generation over the ordered case for suitably chosen system parameters -- disorder-induced order;  (2)  increment in the interaction strength of the disordered case leads to a more increase in the power  than that of the ordered one. It implies that the curvature of quenched averaged power, \(\langle P_{max}\rangle\), in the model with random interactions has  sharper  increase  towards the maximum than the system without any impurities. 

\begin{figure}
\includegraphics[width=\linewidth]{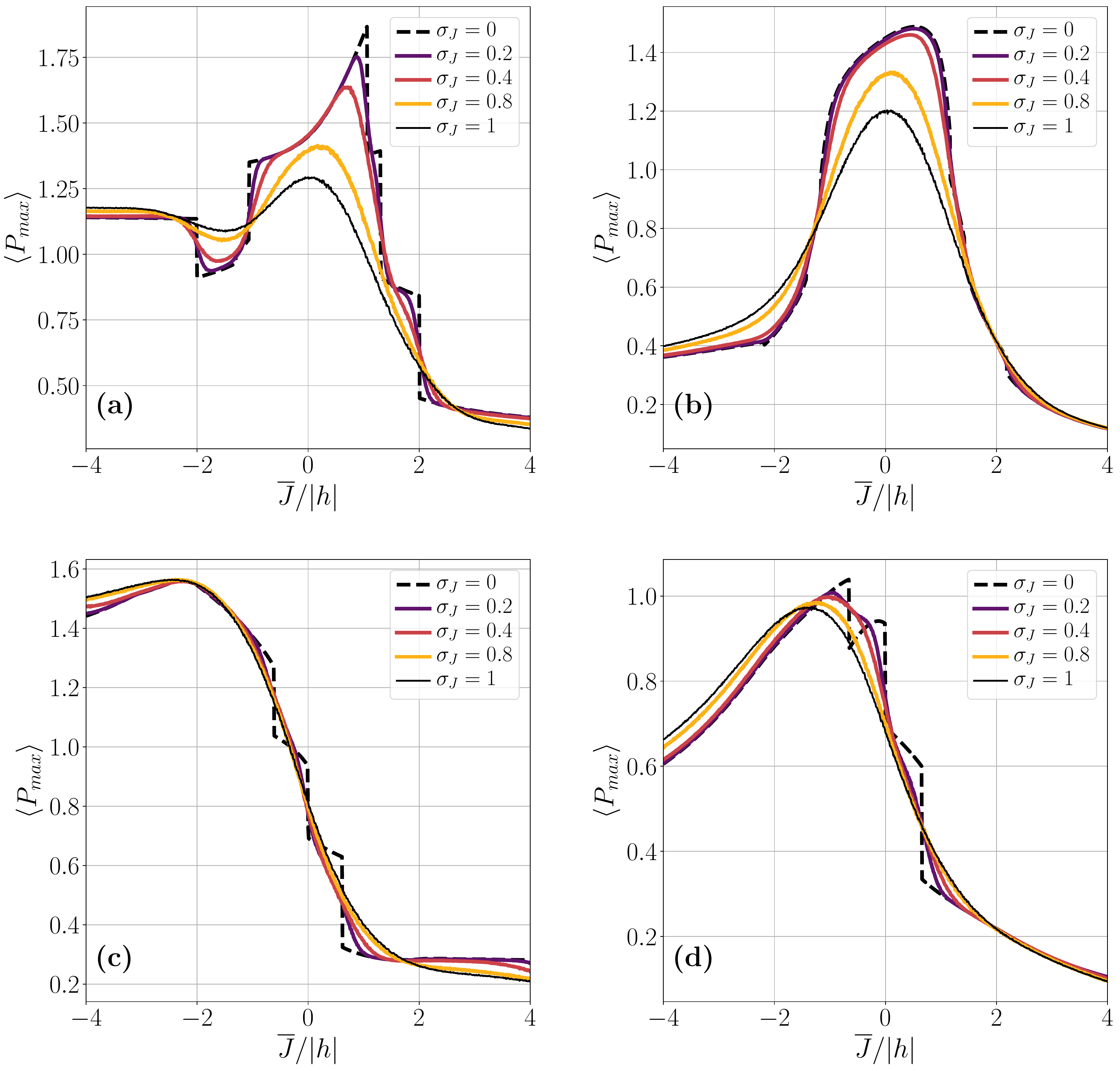}
\caption{(Color online.) Quenched averaged power, $\langle P_{max} \rangle$, against \(\overline{J}/|h|\) for different disorder strength \(\sigma_J\). Note that \(\sigma_J =0\) refers to the ordered case. Disorder  is introduced in the coupling constant  in the $xy$-plane, $J_j/|h|$, for  fixed values of \(\Delta/|h|\) and \(\gamma\).   
The choices of \(\Delta/|h|\) and \(\gamma\) are same as in Fig. \ref{fig:therm2}. The twin advantages mentioned in the text can be visualized from the plots with  \(\Delta/|h| \neq 0\).  Both the axes are dimensionless. }
\label{fig:disJ} 
\end{figure}

\subsection{Effects of  Randomness in $XY$-exchange interaction}

Let us concentrate on the first scenario with  $\{\Delta_j/|h|\}=\Delta/|h|$ and  the disorder being in $\{J_j/|h|\}$, chosen from the Gaussian distribution with a given mean, $\overline{J}/|h|$, and a standard deviation, $\sigma_J$. As mentioned in Sec. \ref{subsec_disorderQXYZ} to obtain the quenched averaged value of the power,  we  here perform averaging over  $5000$  realizations, which we find to be sufficient to converge \(\langle P_{max}\rangle\) upto a second decimal place. 
 Below we emphasize our primary observations regarding the effects of randomness in $XY$-couplings as depicted in Fig. \ref{fig:disJ}.

\begin{enumerate}
\item  For $\Delta/|h|=0$, i.e., for the transverse $XY$ model, increasing the mean interaction strength, $|\overline{J}/h|$,  from $\overline{J}/|h|=0$, does not help to increase the maximum power over the ordered scenario (Fig. \ref{fig:disJ} (a)-(b)). On the other hand, for given values of system parameters, there are situations, both in $\overline{J}/|h|>0$ and $\overline{J}/|h|<0$ -regions, where increasing disorder strength, \(\sigma_J\),  results better production of power, \(\langle P_{max} \rangle \), than that in the ordered case, thereby showing disorder-induced power output. Such advantages is  prominent for lower values of the anisotropy parameter, $\gamma$, and negative values of $\overline{J}/|h|$ (Fig. \ref{fig:disJ} (a)-(b)).

\item Interestingly, in presence of strong and constant interaction in the $z$-direction (e.g.,  when $\Delta/|h|=1$ as shown in Fig. \ref{fig:disJ} (c) and (d)), we find that  for $\overline{J}/|h|<0$, there are situations where we can get better quenched averaged power output by increasing $|\overline{J}/|h|$ than the one obtained in the ordered $XYZ$ model. Secondly, for fixed values of system parameters, $|\overline{J}/h|$, battery produces more power with the increase of  \(\sigma_J\).  Specifically, we observe that  there exists regions in   $|\overline{J}/h|$ where $\langle P_{max} \rangle$ with \(\sigma_J=1\) produces maximum power than any values of \(\sigma_J\). Moreover, as shown in all the situations, increase in the anisotropy parameter suppresses the power generation from the battery. 

\end{enumerate}

\begin{figure}
\includegraphics[width=\linewidth]{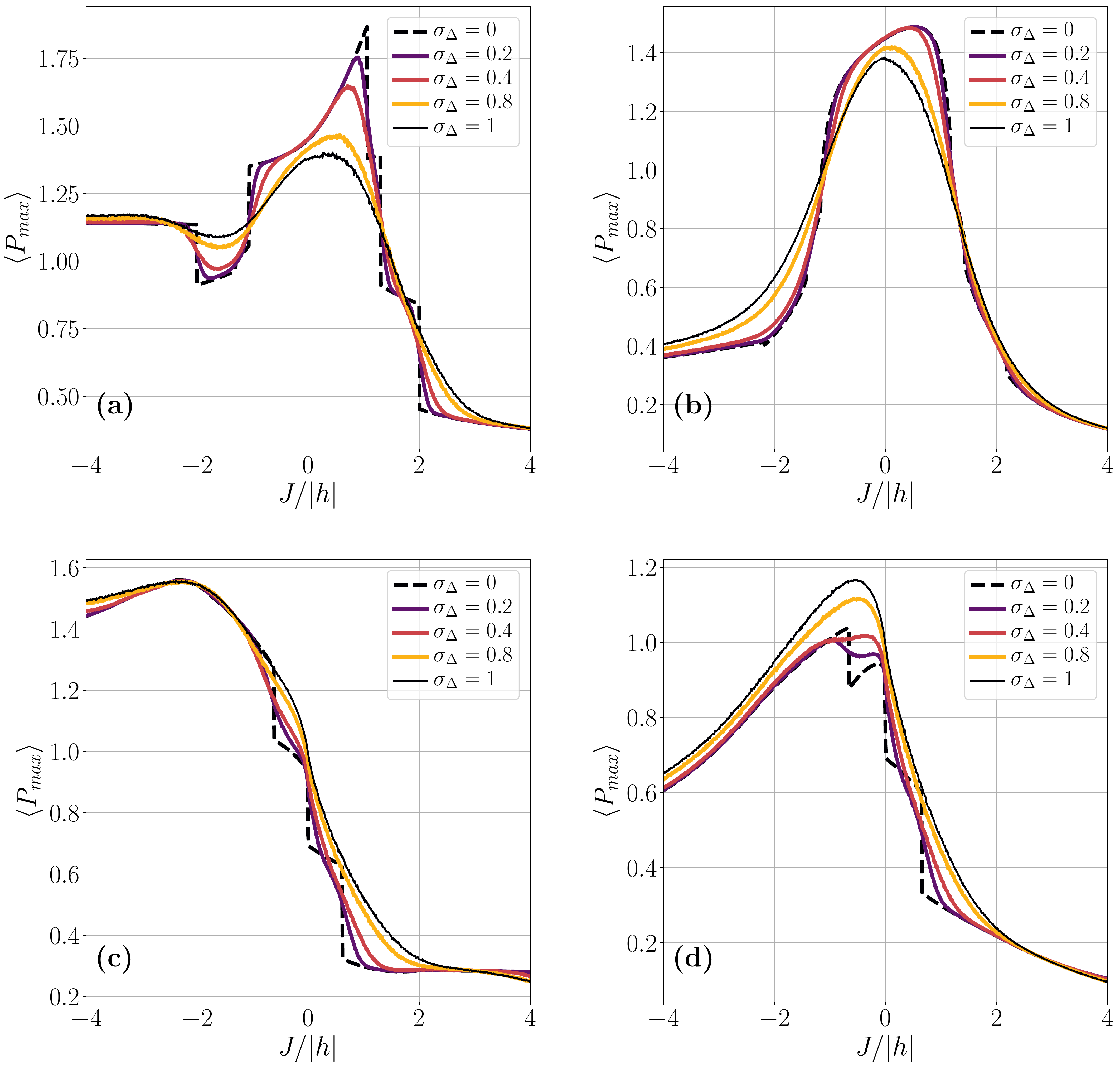}
\caption{(Color online.) \(\langle P_{max}\rangle\) with \(J/|h|\) for  specific choices of mean $\overline{\Delta}/|h|$ and the anisotropy parameter, $\gamma$.  Plots show the effects of disorder in the interaction, $\{\Delta_j\}$, in the $z$-direction, on the  power for different disorder strength, $\sigma_{\Delta}$.  The choices of \(\overline{\Delta}/|h|\) and \(\gamma\) are same as in Fig. \ref{fig:therm2}. Both the axes are dimensionless. }
\label{fig:disDelta} 
\end{figure}

\subsection{Effects of Impurities in the interaction strength in $z$-direction}

Let us now move to the case where randomness is introduced in the interaction strength in the  $z$-direction, i.e., $\{\Delta_j/|h|\}$ are taken randomly from Gaussian distribution with mean, $\overline{\Delta}/|h|$, and standard deviation, $\sigma_{\Delta}$, with keeping $\{J_j/|h|\} = J/|h|$ fixed for every sites (Fig. \ref{fig:disDelta}). As before, we take $5000$ different realizations for quenching.

Comparing Figs. \ref{fig:disDelta} (a) - (b) with \ref{fig:disJ} (a) - (b), we safely claim that   the pattern of  \(\langle P_{max} \rangle \) for model with $\overline{\Delta}/|h| =0\) is almost identical to the  disordered transverse $XY$ model. Note that $\overline{\Delta}/|h| =0\) refers to the disordered $XYZ$ model and does not correspond to the $XY$ model. 

However,  it turns out that the Hamiltonian with $\overline{\Delta}/|h| >0$ is much more beneficial (see Fig. \ref{fig:disDelta}) as compared to the previous cases, where   randomness was in $\{J_j/|h|\}$ and also when  $\overline{\Delta}/|h| =0\). Two prominent differences between these two types of disordered scenarios are  as follows:

\begin{enumerate}

\item Advantages in power with increasing disorder strength and fixed values of system parameters are  less affected by increasing $\gamma$ than any previous situations considered in this paper. Instead of diminishing  the power, we find that the moderate values of \(\gamma\) leads to more efficiency in  power production of the battery in presence of strong disorder. 
 
 \item With non-zero $\overline{\Delta}/|h|$,  we observe that the  quenched averaged power increases with the variation of  \(\sigma_\Delta \) for the entire region of \(|J/h|\), thereby showing advantages of systems having impurities for preparing quantum battery. In particular, as seen 
 in Fig. \ref{fig:disDelta}(d) with $\overline{\Delta}/|h| = 1$ and \(\gamma =0.4\), \(\sigma_\Delta =1\) generates maximum quenched power, \(\langle P_{max} \rangle\) than any other values of \(\sigma_\Delta\). As argued before for the $XY$ and the $XYZ$ models, this kind of advantage  can also be explained as follows:  We choose \(\bar{\Delta}/|h| =1\), from a Gaussian distribution with mean unity and standard deviation \(\sigma_{\Delta}\) which implies  that the   value of \(\Delta/|h|\)  are  approximately  between  \(1- 3 \sigma_{\Delta}\) and  \(1 + 3 \sigma_{\Delta}\).  Thus the nonvanishing nearest-neighbor interaction along with the magnetic field in the \(z\)-direction dominates over the \(xy\)-coupling which is not possible in the $XY$ model and hence the driving field in the $x$-direction requires more energy to take out the system from equilibrium, thereby producing more power. 
 Such a phenomenon  of having advantage of disordered system over the clean case can be referred as disorder-induced order observed in dynamics. \\

\end{enumerate}

\section{Conclusion}
\label{sec-conclu}
Batteries convert chemical energy to the electrical one, thereby
accomplishing our high demands of electricity in daily life. On the other
hand, technological developments lead to the devices which is smaller and
smaller in size, and hence the effects of quantum mechanics on them are
inevitable. Moreover, it was discovered that quantum-based technologies
are more efficient than the existing classical ones. Therefore, it is
natural to explore whether storage devices can also be improved by using
quantum mechanics. It was recently found that this is indeed the case.

If we build quantum battery which is initially prepared in the ground or
thermal states of the quantum spin chain,  the power extracted via local
external magnetic  field
is higher for the interacting models than the noninteracting ones. In
particular, we illustrate the usefulness of interacting Hamiltonian by
considering the ground state of the transverse $XY$  and the $XYZ$ model
with magnetic field as the initial state of the battery. We observe that 
performance of the battery in terms of producing power  declines with the
increase of \(\gamma\). Specifically, the best model which demonstrates
the maximum efficiency is the transverse $XX$ model.  Although the natural
intuition tells us that the  performance of a device can decline  with the
increase of temperature, we find that the suitable tuning of system
parameters leads to a scenario where maximal power generation is higher with the initial state prepared at
 finite temperature than the state with absolute zero-temperature . Finally, we report that impurities help to
improve the generation of quenched averaged power from the battery build up  by using
the ground state of the $XYZ$ model  with random couplings either in the
$xy$-plane or in the $z$-direction in comparison with the ordered systems
--  a phenomena known as disorder-induced order. Both the presence of
impurities and finite temperature are unavoidable  in experiments. Hence
the  enhancement obtained in both the cases indicate that the
implementation of the battery is possible even when the control over the
system is not adequate.

\acknowledgments
The Authors thank Ujjwal Sen for fruitful discussions. 
TC acknowledges support of the National Science Centre (Poland) via QuantERA programme No. 2017/25/Z/ST2/03029.

\section*{APPENDIX A}

 Let us briefly discuss here in details the consequence of finite jumps observed  in Fig. \ref{Fig:powervsJfordifferentgamma;N=8}.  
In this respect, let us first note that finite jumps can only occur for low values of the anisotropy parameter $\gamma$, where interactions in the $x$ and $y$ directions have comparable strengths. Since, we are working with very small systems ($N=4, 6, 8, 10$), the quantum fluctuations are typically large, and the exchange interactions face problems to align (or anti-align) the spins along some specific directions in the $xy$-plane when the value of $\gamma$ is small. That is why the transition points at $J/|h| = \pm 1$ gets bifurcated into different points  which correspond to the finite jumps in the power curve. 

\begin{figure}
\includegraphics[width=\linewidth]{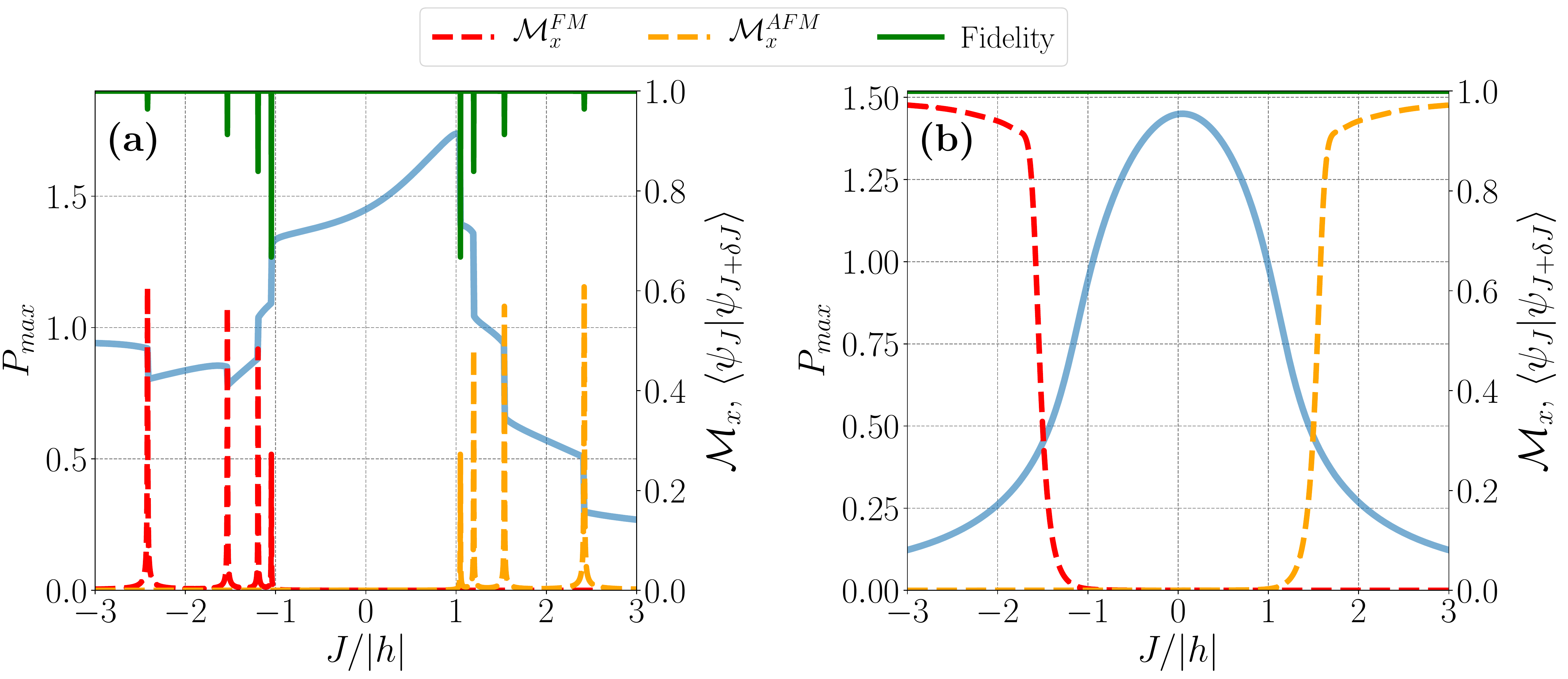}
\caption{(Color online.) Order parameters $\mathcal{M}_x^{FM}$ and $\mathcal{M}_x^{AFM}$, and the fidelity $\braket{\psi_J | \psi_{J+\delta J}}$ for $N=10$ and $\gamma = 0.1$, $0.8$. Here, we take $\delta J = 0.005 |h|$.
The order parameters have non-zero finite values in the corresponding ordered phases, which may not be easily visible in the plots for $\gamma=0.1$.}
\label{fig1}
\end{figure}

We  can confirm this by calculating the ferromagnetic and antiferromagnetic order parameters $$\mathcal{M}_x^{FM} = \sum_j\braket{\sigma_x}/N$$
 and $$\mathcal{M}_x^{AFM} = \sum_j (-1)^j \braket{\sigma_x}/N$$ respectively, as well the fidelity $\braket{\psi_J | \psi_{J+\delta J}}$ (see Fig. \ref{fig1}). Clearly, the finite jumps in power occur exactly at the same positions where the order parameters show non-analyticity and the fidelity shows a dip from $\approx 1$. Hence, we can argue that all the jumps  technically  correspond to the phase transition point, which have been bifurcated from the thermodynamic point due to finite-size effects.  On the other hand, for high values of $\gamma$, say, $0.8$, such a problem cannot persist, since in that case, interaction in the $x$ direction dominates compared to that of  the $y$ direction, and spins can easily align (or anti-align) in the $x$-direction.

\begin{figure}
\includegraphics[width=0.7\linewidth]{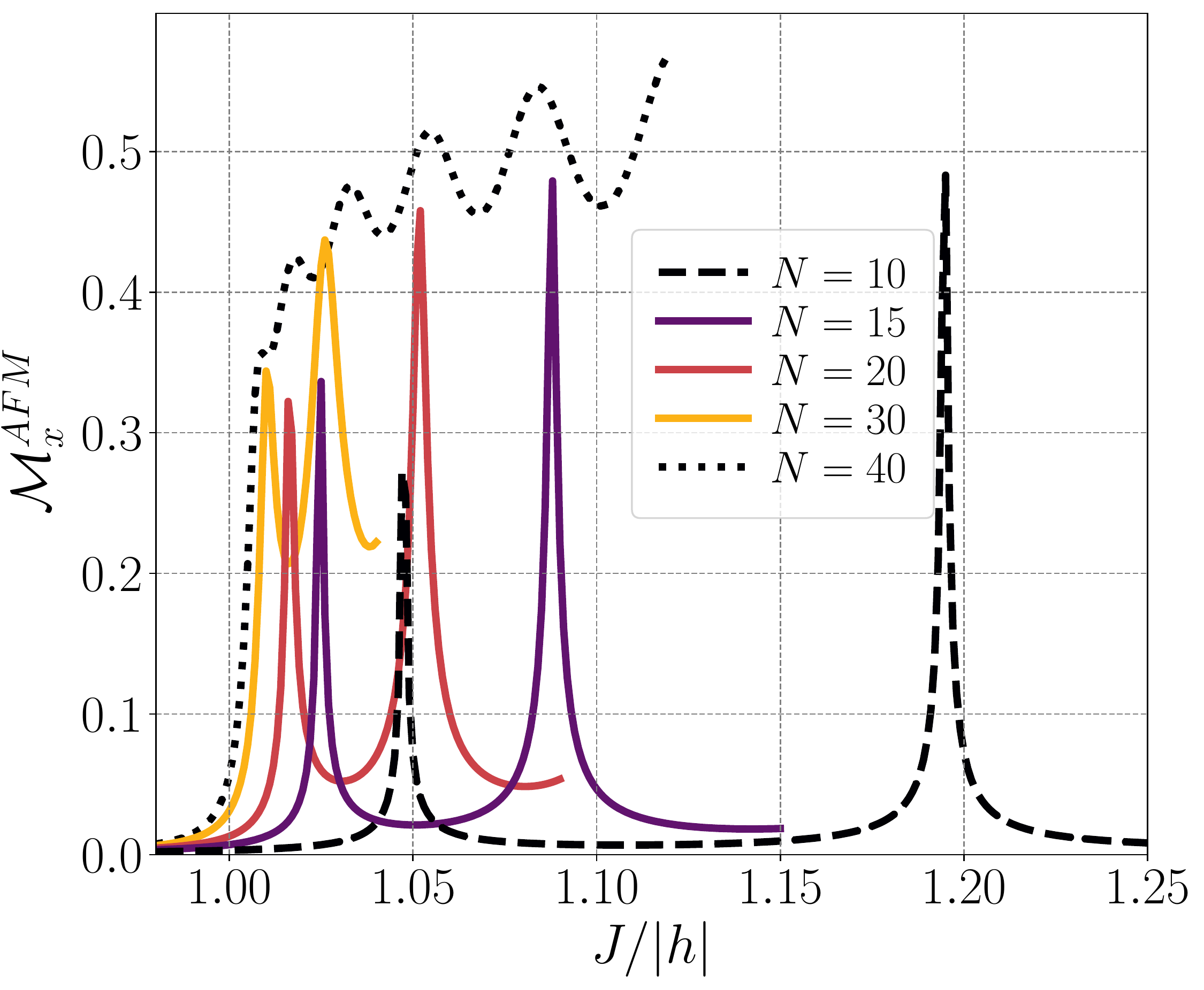}
\caption{(Color online.) Order parameter  $\mathcal{M}_x^{AFM}$ for $\gamma = 0.1$ and $N = 10, 15, 20, 30, 40$.}
\label{fig2}
\end{figure}

To clarify such finite-size effects further, we compute same order parameter  for larger system-sizes. E.g., in Fig. \ref{fig2}, we observe the first two non-analytic points in $\mathcal{M}_x^{AFM}$ for $N=10, 15, 20, 30$, and first few for $N=40$. Clearly, with increasing system-size, all non-analytic points become smoother, and the second one approaches (as well as the later ones also) to the first one and ultimately merges into one. Therefore, we can expect that the first non-analytic point  approaches to the thermodynamic value, i.e., $J/|h|=1$ and hence we consider the first jump in the analysis of scaling. 

For the plots, we employ exact diagonalization method and density matrix renormalization group (DMRG) \cite{dmrg1, dmrg2, dmrg3, dmrg4, dmrg5, dmrg6} technique. For calculating order parameters, we add uniform (or staggered) field of magnitude $10^{-4} |h|$ in the $x$-direction to the Hamiltonian to break the $Z_2$ symmetry.

\end{document}